\documentclass[12pt]{article}

\usepackage{amsmath}

\usepackage{amsfonts}

\usepackage{amssymb}

\usepackage{hyperref}

\usepackage[latin1]{inputenc}

\usepackage{graphics}

\usepackage{graphicx}

\usepackage{epsfig}

\usepackage{amsthm}

\usepackage{mathrsfs}

\usepackage{latexsym}

\usepackage{lineno}

\usepackage{latexsym}

\usepackage{float}

\usepackage[T1]{fontenc}

\usepackage{ae}

\usepackage{authblk}

\modulolinenumbers[5]

\newenvironment{destaque}{\begin{quotation}\small}{\end{quotation}}
\usepackage[figurename=Fig.]{caption}
\usepackage{subcaption}

\newcommand{\h}{\hspace{.5cm}}

\date{}

\begin{document}

\title{\bf An infinite square-well potential as a limiting case of a finite square-well potential in a minimal-length scenario}


\author[1,2]{A. Oakes O. Gon\c{c}alves}
\author[1]{M. F. Gusson}
\author[3]{B. B. Dilem}
\author[4]{R. G. Furtado}
\author[1]{R. O. Francisco}
\author[1]{J. C. Fabris\thanks{julio.fabris@cosmo-ufes.org}}
\author[1]{J. A. Nogueira\thanks{jose.nogueira@ufes.br}}
\affil[1]{Universidade Federal do Esp\'{\i}rito Santo -- Ufes, Vit\' oria, Esp\'{\i}rito Santo, 29.075-910, Brasil}
\affil[2]{Instituto Federal de Alagoas -- Ifal, Batalha, Alagoas, 57.420-000, Brasil}
\affil[3]{Instituto Federal do Esp\'{i}rito Santo -- Ifes, Alegre, Esp\'{\i}rito Santo, 29.520-000, Brasil}
\affil[4]{Universidade Federal do Esp\'{\i}rito Santo -- Ufes, S\~ao Mateus, Esp\'{\i}rito Santo, 29.932-540, Brasil}


\maketitle

\begin{abstract}
\begin{destaque}
 One of the most widely problem studied in quantum mechanics is of an infinite square-well potential. In a minimal-length scenario its study requires additional care because the boundary conditions at the walls of the well are not well fixed. In order to avoid this we solve the finite square-well potential whose the boundary conditions are well fixed, even in a minimal-length scenario, and then we take the limit of the potential going to infinity to find the eigenfunctions and the energy equation for the infinite square-well potential. Although the first correction for the energy eigenvalues is the same one has been found in the literature, our result shows that the eigenfunctions have the first derivative continuous at the square-well walls what is in disagreement those previous work. That is because in the literature the authors  have neglected the hyperbolic solutions and have assumed the discontinuity of the first derivative of the eigenfunctions at the walls of the infinite square-well which is not correct. As we show, the continuity of the first derivative of the eigenfunctions at the square-well walls guarantees the continuity of the probability current density and the unitarity of the time evolution operator.  
\end{destaque}
\end{abstract}

\pagenumbering{arabic}


\section{Introduction}
\label{introd}

\h Although most of the  proposed beyond-Standard-Model theories which try incorporate the gravity lead to existence of a minimal length \cite{Brostein,Mead,Amati}, that is, the existence of a length scale in which the sense of distance loses its meaning, experimental evidences of its actual existence  are very difficult to be accomplished. If the value of the minimal length is of order of the Planck scale, $10^{-35}$ m, the implemention of experiments which confirm its existence may be very far from the current technologies. However, extra dimensions theories suggest that the scale of that minimal length can be many magnitudes greater than the Planck scale \cite{arkani,RS,appelquist,Hossenfelder2}.

A minimal-length scenario can be accomplished in quantum theory by imposing a non-zero minimal uncertainty in the measurement of position\footnote{Even though it is a natural assumption that a minimal length is described as a non-zero minimal uncertainty in position, this is nontrivial \cite{Mead,Kempf:1994su,Garay,Witten,Adler,Camelia,Calmet,Takeuchi}.} which leads to generalized uncertainty principle (GUP). There are different suggestions of modification of the Heisenberg uncertainty principle (HUP) which implement a minimal-length scenario \cite{Kempf:1994su,ali,pedram,Hassanabadi}. We concern with the most usual of them, proposed by Kempf \cite{Kempf:1994su,Kempf2:1997}, which in a 1-dimensional space is given by
\begin{equation}
	\label{gup}
	\Delta x \Delta p \geq \frac{\hbar}{2}  \left[ 1 +  \beta (\Delta p)^2 + \beta \langle \hat{p}  \rangle^{2}  \right],
\end{equation}
where $\beta$ is a parameter related to the minimal length. The GUP (\ref{gup}) implies the existence of a non-zero minimal uncertainty in the position\footnote{That result can easily be obtained from the calculation of the minimum of $\Delta x = f(\Delta p) = \frac{\hbar}{2} \left( \frac{1 + \beta \langle \hat{p}  \rangle^{2}}{\Delta P} + \beta \Delta p \right)$.} $\Delta x_{min} = \hbar \sqrt{\beta}$.
To this generalization of the HUP corresponds to a modification in the canonical commutation relations given by
\begin{equation}
	\label{rc1kempf}
	[\hat{x},\hat{p}] := i\hbar \left(1 + \beta\hat{p}^2 \right).
\end{equation}

In this work, we concern with the finite and the infinite square-well potentials and the more appropriate representation \cite{Brau,pedram2} to be used is\footnote{We also use that representation in order to compare our results with ones of the literature.}
\begin{equation}
\label{xrep}
\hat{x} = \hat{x}_{o}
\end{equation}
and
\begin{equation}
\label{prep}
\hat{p} = \frac{\tan \left( \sqrt{\beta}\hat{p}_{o} \right)}{\sqrt{\beta}},
\end{equation}
where $\hat{x}_{o}$ and $\hat{p}_{o}$ are ordinary operators of position and momentum satisfying the canonical commutation relation $[\hat{x}_{o},\hat{p}_{o}] = i \hbar$. It is not difficult to verify that the (3) and (4) representations for the position and the momentum operators satisfy the (2) commutation relation.
Consequently, to first-order in $\beta$ parameter, we have
\begin{equation}
\label{qp-repx}
\langle x | \hat{x} | \psi(t) \rangle = x \psi (x,t),
\end{equation}
\begin{equation}
\label{qp-repp}
\langle x | \hat{p} | \psi(t) \rangle = -i \hbar \left( 1 - \beta \frac{\hbar^{2}}{3} \frac{\partial^{2}}{\partial x^{2}} \right) \frac{\partial \psi (x,t)}{\partial x},
\end{equation}

The time-independent Schroedinger equation for a non-relativistic particle of mass $m$ up to ${\cal O}(\beta)$ in this representation becomes
\begin{equation}
\label{SchrEq}
-\frac{\hbar^{2}}{2m}\frac{d^{2} \varphi(x)}{dx^{2}} + \beta \frac{\hbar^{4}}{3m}\frac{d^{4} \varphi(x)}{dx^{4}} + V(x)\varphi(x) = E \varphi(x),
\end{equation}
with
\begin{equation}
\label{Psi}
\psi(x,t) = e^{-\frac{i}{\hbar}E t} \varphi(x).
\end{equation}
It follows that the probability density is given by \cite{Vagenas}
\begin{equation}
\label{probd}
\rho = | \psi |^{2}.
\end{equation}
and probability current density by 
$$
J = - \frac{i \hbar}{2m} \left( \psi^{*}\frac{\partial \psi}{\partial x} - \psi \frac{\partial \psi^{*}}{dx} \right) +
$$
\begin{equation}
\label{probc}
\frac{i \beta \hbar^{3}}{m} \left[ \left( \psi^{*}\frac{\partial^{3} \psi}{\partial x^{3}} - \psi \frac{\partial^{3} \psi^{*}}{\partial x^{3}} \right) +
\left( \frac{\partial^{2} \psi^{*}}{\partial x^{2}} \frac{\partial \psi}{\partial x} - \frac{\partial^{2} \psi}{\partial x^{2}} \frac{\partial \psi^{*}}{\partial x} \right) \right].
\end{equation}

The infinite square-well potential is one of the problems most often studied in quantum mechanics. It already shows up in introductory textbooks of quantum physics due to its simplicity and numerous applications. In a minimal-length scenario the study of an infinite square-well requires some special care since many mathematically possible solutions can be found\footnote{Indeed, as we will see, there are four linearly independent solutions for the Schroedinger equation (\ref{SchrEq}).} and the boundary conditions to be imposed on the wave function of the particle and its derivatives at the walls of the infinite square-well are not clearly fixed. Some authors have bypassed that problem neglecting the hyperbolic solutions by stating that they are not physical solutions since they would not satisfy the boundary conditions at the walls of the infinite square-well \cite{nozari,pedram3,pedram4}. But, as it is well-known, the boundary conditions must be satisfied by the general solution, that is, the linear combination of the possible linearly independent solutions. Thus, for only requiring the continuity of the wave function at the walls of the infinite square-well is not possible to determine the solution.

In principle, the continuity or not of the derivatives of the wave function in the wall of a square-well localized at $x = a$ could be obtained by integrating Eq. (\ref{SchrEq}) between $a - \epsilon$ and $a + \epsilon$ (with $\epsilon$ arbitrarily small and positive), and then taking the limit $\epsilon \rightarrow 0$. Thus,
\begin{equation*}
\left[ \frac{d \varphi_{III}(a)}{d x} - \frac{d \varphi_{II}(a)}{d x} \right] - 
\frac{2}{3}\beta \hbar^{2} \left[ \frac{d^{3} \varphi_{III}(a)}{d x^{3}} - \frac{d^{3} \varphi_{II}(a)}{d x^{3}} \right] =
\end{equation*}
\begin{equation}
\label{d3d1}
 - \lim_{\epsilon \rightarrow 0} \frac{2m}{\hbar^{2}}\int_{a -\epsilon}^{a +\epsilon} dx V(x)\varphi(x),
\end{equation}
where $\varphi_{II}(x)$ and $\varphi_{III}(x)$ are the solutions of Eq. (\ref{SchrEq}) for $x < a$ and $x > a$, respectively.

Since $\varphi(x)$ must be everywhere finite, if $V(x)$ has a finite jump at $x = a$ then the fourth derivative of $\varphi(x)$ at $x = a $ has a finite discontinuity (that is to say, a jump by a finite amount), so we require that the third, second and first derivatives are continuous\footnote{That occurs because any function always has an infinite first derivative at point which it is discontinuous. Therefore in order the fourth derivative of a function to be finite it is necessary that its third derivative be continuous \cite{Eisberg}.} at $x = a$. However, if  $V(x)$ has an infinite jump at $x = a$ then the third and/or first derivative of $\varphi(x)$ has a discontinuity at $x = a$. Hence, we have an ignorance of the behavior of the first derivative at $x = a$. Therefore, for assuming the discontinuity of the first derivative can be a mistaken assumption which will lead to incorrect results. A way to get around this problem is to solve the finite square-well potential, whose the boundary conditions are well fixed, and then to take the limit of the potential going to infinity\footnote{We have already that difficulty in ordinary quantum mechanics, because the Schroedinger equation does not provide us a guide on how to fix the boundary conditions for the case of an infinite square-well potential \cite{Griffiths,McIntyre,Robinett}. In 1938, Rojansky pointed out that this difficulty can be solved in a mathematically rigorous way by considering an infinite square-well as the limit case of a finite square-well \cite{Rojansky,Seki,Bonneau}.}. This way we can check the results have been found in the literature \cite{nozari,pedram3,pedram4,Esguerra}. With this goal in mind, we solve a symmetric finite square-well in a minimal-length scenario in Section \ref{Ddfp} (in fact, the calculations can be found in the Appendix) and, in Section \ref{Ddip}, we take the limit of the potential going to infinity in order to find the solution for an infinite square-well in that scenario. As we will see, the result for the eigenfunctions found is in disagreement with ones of the literature whereas the first correction for the energy eigenvalues is the same.


\section{Finite square-well potential}
\label{Ddfp}

\h We will consider that the potential $V(x)$ of Eq. (\ref{SchrEq}) be given by 
\begin{equation}
\label{fsw} 
    V(x) =
	\left \{
	\begin{array}{ll}
	V_{0}, & |x| \geq \frac{a}{2} \\ [11pt]
	0, & -\frac{a}{2} < x < \frac{a}{2}, \\
	\end{array}
	\right.
\end{equation}
where $V_{0}$ is a real and positive constant. We concern with the case $E < V_{0}$ since there are only bound states in the infinite square-well. Consequently, the general solution of Eq. (\ref{SchrEq}) is given by\footnote{We have thrown away those terms which diverge when $x \rightarrow \pm \infty$.} 
\begin{equation}
\label{gensol} 
	\left \{
	\begin{array}{ll}
	\varphi_{I}(x) = A_{1}e^{k_{1}^{\beta} x} + C_{1} e^{k^{-}_{\beta} x}  & x \leq -\frac{a}{2} \\ [11pt]
	\varphi_{II}(x) = A_{2}e^{-ik_{0}^{\beta} x} + B_{2}e^{ik_{0}^{\beta} x} + C_{2} e^{-k^{+}_{\beta} x} + D_{2} e^{k^{+}_{\beta} x}, & -\frac{a}{2} \leq x \leq \frac{a}{2}, \\  [11pt]
	\varphi_{III}(x) = A_{3}e^{-k_{1}^{\beta} x} + C_{3} e^{-k^{-}_{\beta} x}  & x \geq \frac{a}{2} \\
	\end{array}
	\right.
\end{equation}
where\footnote{Note that the $E$ energy has corrections in orders of $\beta$.}
\begin{equation}
\label{k0}
k_{0} := \sqrt{\frac{2mE}{\hbar^{2}}}, 
\end{equation}
\begin{equation}
\label{k1}
k_{1} := \sqrt{\frac{2m \left( V_{0} - E \right)}{\hbar^{2}}}, 
\end{equation}
\begin{equation}
\label{k0b}
k_{0}^{\beta}:= k_{0} \left( 1 - \frac{1}{3} \beta \hbar^{2} k_{0}^{2} \right)  + {\cal O}(\beta^{2}), 
\end{equation}
\begin{equation}
\label{k1b}
k_{1}^{\beta} := k_{1} \left( 1 + \frac{1}{3} \beta \hbar^{2} k_{1}^{2} \right)  + {\cal O}(\beta^{2}), 
\end{equation}
\begin{equation}
\label{kbetamais}
k_{\beta}^{+} := \sqrt{\frac{3}{2 \hbar^{2} \beta}} \left( 1 + \frac{1}{3} \beta \hbar^{2} k_{0}^{2} \right)  + {\cal O}(\beta^{3/2}),
\end{equation}
\begin{equation}
\label{kbetamenos}
k_{\beta}^{-} := \sqrt{\frac{3}{2 \hbar^{2} \beta}} \left( 1 - \frac{1}{3} \beta \hbar^{2} k_{1}^{2} \right)  + {\cal O}(\beta^{3/2}).
\end{equation}

As was said before, since $\varphi(x)$ and $V_{0}$ are finite the fourth derivative of $\varphi(x)$ has a finite discontinuity jump at $x = \pm \frac{a}{2}$ which implies that the third derivative of $\varphi(x)$ is continuous at $x = \pm \frac{a}{2}$. Consequently, the second and first derivatives of $\varphi(x)$ and $\varphi(x)$ are continuous at $x = \pm \frac{a}{2}$, too. Thus \cite{Vagenas1},

\begin{equation}
\label{bc1}
\begin{array}{ll}
\frac{d^{n} \varphi_{I}(x)}{dx^{n}} \bigg|_{x = -\frac{a}{2}} = \frac{d^{n} \varphi_{II}(x)}{dx^{n}} \bigg|_{x = -\frac{a}{2}}, &  n = 0, 1, 2, 3,
\end{array}
\end{equation}
and
\begin{equation}
\label{bc2}
\begin{array}{ll}
\frac{d^{n} \varphi_{II}(x)}{dx^{n}} \bigg|_{x = \frac{a}{2}} = \frac{d^{n} \varphi_{III}(x)}{dx^{n}} \bigg|_{x = \frac{a}{2}},  &  n = 0, 1, 2, 3.
\end{array}
\end{equation}

These boundary conditions are in accordance with the physical conditions of the probability density, Eq. (\ref{probd}), and of the probability current density, Eq. (\ref{probc}), must be everywhere continuous for a finite potential \cite{Ballentine}.

The application of the above boundary conditions leads to a system of equations whose 8 unknowns are the coefficients of Eq. (\ref{gensol}). The calculation can be simplified if we note that the potential (\ref{fsw}) is even, that is, $V(-x) = V(x)$. This way we have even solutions for which $\varphi(-x) = \varphi(x)$, and odd solutions for which $\varphi(-x) = -\varphi(x)$.

\subsection{Even solutions}
\label{Esolsec}

\h In this case, $A_{1} = A_{3}$, $C_{1} = C_{3}$, $A_{2} = B_{2}$, $C_{2} = D_{2}$, and the solution is
\begin{equation}
\label{esol} 
	\left \{
	\begin{array}{ll}
	\varphi_{I}(x) = A_{3}e^{k_{1}^{\beta} x} + C_{3} e^{k^{-}_{\beta} x}  & x \leq -\frac{a}{2} \\ [11pt]
	\varphi_{II}(x) = A_{2}\cos \left( k_{0}^{\beta} x \right) + C_{2} \cosh \left( k^{+}_{\beta} x \right), & -\frac{a}{2} \leq x \leq \frac{a}{2}, \\ [11pt]
	\varphi_{III}(x) = A_{3}e^{-k_{1}^{\beta} x} + C_{3} e^{-k^{-}_{\beta} x}  & x \geq \frac{a}{2} \\
	\end{array}
	\right.
\end{equation}
where (see \ref{sle})
\begin{equation}
\label{A3p}
A_{3} = \frac{ \left( k^{-}_{\beta} \right)^{2} + \left( k_{0}^{\beta} \right)^{2}}{\left( k^{-}_{\beta} \right)^{2}  - \left( k_{1}^{\beta} \right)^{2}} \left[ \frac{ k_{0}^{\beta}\tan \left( \frac{ k_{0}^{\beta} a}{2} \right) + k_{\beta}^{+}\tanh \left( \frac{ k_{\beta}^{+} a}{2} \right)}{ k_{1}^{\beta} + k_{\beta}^{+}\tanh \left( \frac{ k_{\beta}^{+} a}{2} \right)} \right] \cos \left( \frac{ k_{0}^{\beta} a}{2} \right) \exp\left( {\frac{k_{1}^{\beta} a}{2}} \right) A_{2},
\end{equation}
\\
\begin{equation}
\label{C2p}
C_{2} = - \frac{ \left( k^{-}_{\beta} \right)^{2} + \left( k_{0}^{\beta} \right)^{2}}{\left( k^{-}_{\beta} \right)^{2}  - \left( k_{\beta}^{+} \right)^{2}} \left[ \frac{ k_{1}^{\beta} - k_{0}^{\beta}\tan \left( \frac{ k_{0}^{\beta} a}{2} \right)}{ k_{1}^{\beta} + k_{\beta}^{+} \tanh \left( \frac{ k_{\beta}^{+} a}{2} \right)} \right] \frac{\cos \left( \frac{ k_{0}^{\beta} a}{2} \right)} {\cosh \left( \frac{ k_{\beta}^{+} a}{2} \right)} A_{2},
\end{equation}
and
\\
\begin{equation*}
C_{3} = \Bigg\{ 1 - \frac{ \left( k^{-}_{\beta} \right)^{2} + \left( k_{0}^{\beta} \right)^{2}}{\left( k^{-}_{\beta} \right)^{2}  - \left( k_{1}^{\beta} \right)^{2}} \left[ \frac{ k_{0}^{\beta}\tan \left( \frac{ k_{0}^{\beta} a}{2} \right) + k_{\beta}^{+}\tanh \left( \frac{ k_{\beta}^{+} a}{2} \right)}{ k_{1}^{\beta} + k_{\beta}^{+}\tanh \left( \frac{ k_{\beta}^{+} a}{2} \right)} \right] -
\end{equation*}
\begin{equation}
\label{C3p}
\frac{ \left( k^{-}_{\beta} \right)^{2} + \left( k_{0}^{\beta} \right)^{2}}{\left( k^{-}_{\beta} \right)^{2}  - \left( k_{\beta}^{+} \right)^{2}} \left[ \frac{ k_{1}^{\beta} - k_{0}^{\beta}\tan \left( \frac{ k_{0}^{\beta} a}{2} \right)}{ k_{1}^{\beta} + k_{\beta}^{+} \tanh \left( \frac{ k_{\beta}^{+} a}{2} \right)} \right] \Bigg\} \cos \left( \frac{ k_{0}^{\beta} a}{2} \right) \exp \left( {\frac{k_{\beta}^{-}a}{2}} \right) A_{2}.
\end{equation}

Furthermore, we also obtain the energy equation:
\begin{equation*}
k_{0}^{\beta}\tan \left( \frac{ k_{0}^{\beta} a}{2} \right) - k_{\beta}^{-} +
\end{equation*}
\begin{equation*}
\frac{ \left( k^{-}_{\beta} \right)^{2} + \left( k_{0}^{\beta} \right)^{2}}{\left( k^{-}_{\beta} \right)^{2}  - \left( k_{\beta}^{+} \right)^{2}}\left[ \frac{ k_{1}^{\beta} - k_{0}^{\beta}\tan \left( \frac{ k_{0}^{\beta} a}{2} \right)}{ k_{1}^{\beta} + k_{\beta}^{+} \tanh \left( \frac{ k_{\beta}^{+} a}{2} \right)} \right]\left[k_{\beta}^{-} + k_{\beta}^{+} \tanh \left( \frac{ k_{\beta}^{+} a}{2} \right) \right] +
\end{equation*}
\begin{equation}
\label{eeeq}
\frac{ \left( k^{-}_{\beta} \right)^{2} + \left( k_{0}^{\beta} \right)^{2}}{\left( k^{-}_{\beta} \right)^{2}  - \left( k_{1}^{\beta} \right)^{2}}\left[ \frac{ k_{0}^{\beta}\tan \left( \frac{ k_{0}^{\beta} a}{2} \right) + k_{\beta}^{+}\tanh \left( \frac{ k_{\beta}^{+} a}{2} \right)}{ k_{1}^{\beta} + k_{\beta}^{+}\tanh \left( \frac{ k_{\beta}^{+} a}{2} \right)} \right]\left( k_{\beta}^{-} - k_{1}^{\beta} \right) = 0.
\end{equation}

In the limit when $\beta$ goes to zero ($\beta \rightarrow 0$) the solutions (\ref{esol}) and the energy equation (\ref{eeeq}) become the even solutions and the energy equation, respectively, for the finite square-well potential in ordinary quantum mechanics, as we expected.

\subsection{Odd solutions}
\label{Osolsec}

\h In this case, $A_{1} = -A_{3}$, $C_{1} = -C_{3}$, $A_{2} = -B_{2}$, $C_{2} = -D_{2}$, and the solution is
\begin{equation}
\label{osol} 
	\left \{
	\begin{array}{ll}
	\varphi_{I}(x) = -A_{3}e^{k_{1}^{\beta} x} - C_{3} e^{k^{-}_{\beta} x}  & x \leq -\frac{a}{2} \\ [11pt]
	\varphi_{II}(x) = B_{2}\sin \left( k_{0}^{\beta} x \right) + D_{2} \sinh \left( k^{+}_{\beta} x \right), & -\frac{a}{2} \leq x \leq \frac{a}{2}, \\ [11pt]
	\varphi_{III}(x) = A_{3}e^{-k_{1}^{\beta} x} + C_{3} e^{-k^{-}_{\beta} x}  & x \geq -\frac{a}{2} \\
	\end{array}
	\right.
\end{equation}
where
\begin{equation}
\label{A3i}
A_{3} = \frac{ \left( k^{-}_{\beta} \right)^{2} + \left( k_{0}^{\beta} \right)^{2}}{\left( k^{-}_{\beta} \right)^{2}  - \left( k_{1}^{\beta} \right)^{2}} \left[ \frac{ k_{\beta}^{+} \coth \left( \frac{ k_{\beta}^{+} a}{2} \right) - k_{0}^{\beta}\cot \left( \frac{ k_{0}^{\beta} a}{2} \right) }{ k_{1}^{\beta} + k_{\beta}^{+}\coth \left( \frac{ k_{\beta}^{+} a}{2} \right)} \right] \sin \left( \frac{ k_{0}^{\beta} a}{2} \right) \exp \left( \frac{k_{1}^{\beta} a}{2} \right) B_{2},
\end{equation}
\\
\begin{equation}
\label{d2i}
D_{2} = - \frac{ \left( k^{-}_{\beta} \right)^{2} + \left( k_{0}^{\beta} \right)^{2}}{\left( k^{-}_{\beta} \right)^{2}  - \left( k_{\beta}^{+} \right)^{2}} \left[ \frac{ k_{1}^{\beta} + k_{0}^{\beta}\cot \left( \frac{ k_{0}^{\beta} a}{2} \right)}{ k_{1}^{\beta} + k_{\beta}^{+} \coth \left( \frac{ k_{\beta}^{+} a}{2} \right)} \right] \frac{\sin \left( \frac{ k_{0}^{\beta} a}{2} \right)} {\sinh \left( \frac{ k_{\beta}^{+} a}{2} \right)} B_{2},
\end{equation}
and
\begin{equation*}
C_{3} = \Bigg\{ 1 - \frac{ \left( k^{-}_{\beta} \right)^{2} + \left( k_{0}^{\beta} \right)^{2}}{\left( k^{-}_{\beta} \right)^{2}  - \left( k_{1}^{\beta} \right)^{2}} \left[ \frac{ k_{\beta}^{+}\coth \left( \frac{ k_{\beta}^{+} a}{2} \right) - k_{0}^{\beta}\cot \left( \frac{ k_{0}^{\beta} a}{2} \right) }{ k_{1}^{\beta} + k_{\beta}^{+}\coth \left( \frac{ k_{\beta}^{+} a}{2} \right)} \right] -
\end{equation*}
\begin{equation}
\label{C3i}
\frac{ \left( k^{-}_{\beta} \right)^{2} + \left( k_{0}^{\beta} \right)^{2}}{\left( k^{-}_{\beta} \right)^{2}  - \left( k_{\beta}^{+} \right)^{2}} \left[ \frac{ k_{1}^{\beta} + k_{0}^{\beta}\cot \left( \frac{ k_{0}^{\beta} a}{2} \right)}{ k_{1}^{\beta} + k_{\beta}^{+} \coth \left( \frac{ k_{\beta}^{+} a}{2} \right)} \right] \Bigg\} \sin \left( \frac{ k_{0}^{\beta} a}{2} \right) \exp \left( {\frac{k_{\beta}^{-}a}{2}} \right) B_{2}.
\end{equation}

Furthermore, we also obtain the energy equation:
\begin{equation*}
k_{0}^{\beta}\cot \left( \frac{ k_{0}^{\beta} a}{2} \right) + k_{\beta}^{-} -
\end{equation*}
\begin{equation*}
\frac{ \left( k^{-}_{\beta} \right)^{2} + \left( k_{0}^{\beta} \right)^{2}}{\left( k^{-}_{\beta} \right)^{2}  - \left( k_{\beta}^{+} \right)^{2}}\left[ \frac{ k_{1}^{\beta} + k_{0}^{\beta}\cot \left( \frac{ k_{0}^{\beta} a}{2} \right)}{ k_{1}^{\beta} + k_{\beta}^{+} \coth \left( \frac{ k_{\beta}^{+} a}{2} \right)} \right]\left[k_{\beta}^{-} + k_{\beta}^{+} \coth \left( \frac{ k_{\beta}^{+} a}{2} \right) \right] +
\end{equation*}
\begin{equation}
\label{oeeq}
\frac{ \left( k^{-}_{\beta} \right)^{2} + \left( k_{0}^{\beta} \right)^{2}}{\left( k^{-}_{\beta} \right)^{2}  - \left( k_{1}^{\beta} \right)^{2}}\left[ \frac{ k_{\beta}^{+}\coth \left( \frac{ k_{\beta}^{+} a}{2} \right) - k_{0}^{\beta}\cot \left( \frac{ k_{0}^{\beta} a}{2} \right)}{ k_{1}^{\beta} + k_{\beta}^{+}\coth \left( \frac{ k_{\beta}^{+} a}{2} \right)} \right]\left( k_{1}^{\beta} - k_{\beta}^{-} \right) = 0.
\end{equation}

Again, in the limit when $\beta$ goes to zero ($\beta \rightarrow 0$) the solutions (\ref{osol}) and the energy equation (\ref{oeeq}) become the odd solutions and the energy equation, respectively, for the finite square-well potential in ordinary quantum mechanics, as we also expected.


\section{Infinite square-well potential}
\label{Ddip}

\h  As previously mentioned, the energy eigenfunctions and the energy equation of the infinite square-well can be obtained from the finite square-well taking the limit $V_{0} \rightarrow \infty$. In this limit $A_{3} e^{- k_{1}^{\beta}|x|} = 0$, $C_{3}e^{- k_{\beta}^{-} |x|} = 0$, and 
\begin{equation}
\label{iC2p}
C_{2} = - \frac{ \cos \left( \frac{ k^{\beta}_{0} a}{2} \right)}{ \cosh \left( \frac{ k_{\beta}^{+} a}{2} \right)} A_{2}.
\end{equation}
Therefore, the even solutions, that is, the even eigenfunctions of the infinite square-well are given by
\begin{equation}
\label{iesol} 
	\left \{
	\begin{array}{ll}
	\varphi_{I}(x) = 0 & x \leq -\frac{a}{2} \\ [11pt]
	\varphi_{II}(x) = A_{2}\cos \left( k_{0}^{\beta} x \right) - A_{2}  \frac{\cos \left( \frac{ k_{0}^{\beta} a}{2} \right) }{\cosh \left( \frac{ k_{\beta}^{+} a}{2} \right)} \cosh \left( k^{+}_{\beta} x \right), & -\frac{a}{2} \leq x \leq \frac{a}{2}, \\ [11pt]
	\varphi_{III}(x) =0  & x \geq -\frac{a}{2} \\
	\end{array}
	\right.
\end{equation}
and, the eigenenergies of the even eigenstates are given by the equation
\begin{equation}
\label{ieenergy}
k_{0}^{\beta} \tan \left( \frac{ k_{0}^{\beta} a}{2} \right) + k_{\beta}^{+} \tanh \left( \frac{ k_{\beta}^{+} a}{2} \right) = 0.
\end{equation}
Whereas the coefficients of the odd solutions become $A_{3} =0$, $C_{3} = 0$ and 
\begin{equation}
\label{iC2o}
D_{2} = - \frac{ \sin \left( \frac{ k^{\beta}_{0} a}{2} \right)}{ \sinh \left( \frac{ k_{\beta}^{+} a}{2} \right)} A_{2}.
\end{equation}
Therefore, the odd solutions, that is, the odd eigenfunctions of the infinite square-well are given by
\begin{equation}
\label{iosol} 
	\left \{
	\begin{array}{ll}
	\varphi_{I}(x) = 0 & x \leq -\frac{a}{2} \\ [11pt]
	\varphi_{II}(x) = B_{2}\sin \left( k_{0}^{\beta} x \right) - B_{2}  \frac{\sin \left( \frac{ k_{0}^{\beta} a}{2} \right) }{\sinh \left( \frac{ k_{\beta}^{+} a}{2} \right)} \sinh \left( k^{+}_{\beta} x \right), & -\frac{a}{2} \leq x \leq \frac{a}{2}, \\ [11pt]
	\varphi_{III}(x) =0  & x \geq -\frac{a}{2} \\
	\end{array}
	\right.
\end{equation}
and, the eigenenergies of the odd eigenstates are given by the equation,
\begin{equation}
\label{ioenergy}
k_{0}^{\beta} \cot \left( \frac{ k_{0}^{\beta} a}{2} \right) - k_{\beta}^{+} \coth \left( \frac{ k_{\beta}^{+} a}{2} \right) = 0.
\end{equation}

It is easy to see that in the limit $\beta \rightarrow 0$ we recover the results known for infinite square-well potential in ordinary quantum mechanics\footnote{Note that $\frac{\cosh \left( k^{+}_{\beta} x \right)}{\cosh \left( \frac{k^{+}_{\beta} a}{2} \right)} \rightarrow 0$ and $\frac{\sinh \left( k^{+}_{\beta} x \right)}{\sinh \left( \frac{k^{+}_{\beta} a}{2} \right)} \rightarrow 0$ since $|x| \leq \frac{a}{2}$ for $\varphi_{II}(x)$.}.

It is important to point out that now the first derivative at $x = \pm \frac{a}{2}$ is no longer discontinuous. In contrast what occurs in the ordinary quantum mechanics, in which the continuity of the solutions at $x = \pm \frac{a}{2}$ are obtained by using the energy equation, in a minimal-length scenario the solutions are continuous at $x = \pm \frac{a}{2}$ without need to use the energy equation, which is required for continuity of the first derivative.

The previous results are in disagreement with ones in the literature \cite{nozari,pedram3,pedram4,Esguerra}, in which have been found no change in the eigenfunctions up to the first order in $\beta$. This is because those authors have presumed the discontinuity of the first derivative of the solutions (eigenfunctions) at the walls of the infinite square-well as for ordinary quantum mechanics.

The necessity of continuity of the first derivative of $\varphi(x)$ is evidenced by a quick look at the $J$ probability current density, Eq. (\ref{probc}). In the ordinary case, by requiring that $\varphi \left( x = \pm \frac{a}{2} \right) = 0$ is guaranteed the continuity of the probability current density \cite{Ballentine}. On the other hand, in a minimal-length scenario the continuity condition of $\varphi(x)$ is not enough to guarantee the continuity of the $J$ probability current density at $x = \pm \frac{a}{2}$. However the continuity of $\varphi(x)$ and $\frac{d \varphi}{dx}$, that is, $\varphi \left(\pm \frac{a}{2} \right) = 0$ and $\frac{d \varphi}{dx} \left(\pm \frac{a}{2} \right) = 0$, provides the necessary condition for the continuity of the $J$ probability current density at $x = \pm \frac{a}{2}$.

The continuity of the first derivative does not really come  as a surprise since similar outcome has been found in the study of a Dirac $\delta$-function potential in a minimal-length scenario \cite{Gusson}. Apparently, if corrections of ${\cal O}\left( \beta^{2} \right)$ are taken account third derivative of the wave function becomes continuous, too.

It is important to stress that the energy spectrum is discrete since the eigenvalues of energy are given by points of intersection of the tangent (cotangent) and hyperbolic tangent (hyperbolic cotangent) functions.

In order to find the correction to first-order in $\beta$ of the energy eigenvalues, we write
 $E = E_{0} + \beta \epsilon$, where $E_{0} = \frac{\pi^{2} \hbar^{2} n^{2}}{2 m a^{2}}$, $n = 1, 2, 3, \dots$, and put it in Eqs. (\ref{ieenergy}) and (\ref{ioenergy}). After some algebra we obtain\footnote{Note that $k_{\beta}^{+}$ goes to infinity at $\beta = 0$ while $\tanh \left( \frac{k_{\beta}^{+} a}{2} \right)$ and $\coth \left( \frac{k_{\beta}^{+} a}{2} \right)$ remain finite.}
\begin{equation}
\label{ebeta}
E = \frac{\pi^{2} \hbar^{2} n^{2}}{2 m a^{2}} + \beta \frac{\pi^{4} \hbar^{4} n^{4}}{3 m a^{4}}, 
\end{equation}
with $n = 1, 2, 3, \dots$.
The result Eq. (\ref{ebeta}) is in accordance with the literature \cite{nozari,pedram3,pedram4,Esguerra}. This was expected since the calculate of the first order correction for the energy involves just the non-deformed wave functions.

It easy to see that when the distance between the walls of the well increases more negligible is the correction term of Eq. (\ref{ebeta}) and also are the second terms of the solutions (\ref{iesol}) and (\ref{iosol}) since $\cosh \left( \frac{k^{+}_{\beta} a}{2} \right)$ and $\sinh \left( \frac{k^{+}_{\beta} a}{2} \right)$ increase. However, it is import to stress that from the physical point view it make no sense to take into account cases in which the distance between the walls of the well is equal to or smaller than the minimal length, that is, for cases where $\beta$ is fixed and $\frac{a}{\hbar \sqrt{\beta}} \leq 1$. This is because we would compute probabilities of finding the particle in ranges smaller than the minimal length.  Of course, if $a$ is somewhat greater than the minimal length,  $\frac{a}{\hbar \sqrt{\beta}} \sim 1$, we can calculate probabilities of finding the particle in ranges larger than the minimal length.  But a quick dimensional analyse shows that the correction of order $\beta^{N}$ of $E$, Eq.(\ref{ebeta}),  is of order $\left( \frac{l_{min}}{a} \right)^{2N}$. Then, if $\frac{a}{\hbar \sqrt{\beta}} \sim 1$ the result (\ref{ebeta}) lies far outside the validity range at which we may consistently work. At last, it should point out that even for $\frac{a}{\hbar \sqrt{\beta}}\gg 1$ when $| \frac{a}{2} - x |$ is small enough, the second terms of the solutions (\ref{iesol}) and (\ref{iosol}) become  significant, because of
\begin{equation}
\label{t2se}
\frac{ e^{-k_{\beta}^{+} \left( \frac{a}{2} - x \right)} \pm e^{-k_{\beta}^{+} \left( \frac{a}{2} + x \right)}}{1 \pm e^{-k_{\beta}^{+} a}}.
\end{equation}




\section{Unitarity}
\label{Unitarity}

\h  Since the potential energy in the Schroedinger equation, Eq. (\ref{SchrEq}), is independent of the time, the time evolution operator is given by
\begin{equation}
\label{evop}
\hat{U}(t, t_{0}) = e^{-\frac{i}{\hbar}(t-t_{0})\hat{H}}. 
\end{equation}
Thus, $\hat{U}$ will be unitary if $\hat{H}$ is a Hermitian operator (rigorously, self-adjoint or admits self-adjoint extensions).

The inner product remains the standard one because we have kept the position operator to be usual operator \cite{nozari,Fring}, so the Hamiltonian operator will be Hermitian, $\hat{H}^{\dag} = \hat{H}$, if
\begin{equation}
\label{hbc}
- \left[ \phi^{\star}\frac{d \varphi}{dx} -\varphi\frac{d \phi^{\star}}{dx}  \right]^{b}_{a} + \beta \left[ \phi^{\star}\frac{d^{3} \varphi}{dx^{3}} -\varphi\frac{d^{3} \phi^{\star}}{dx^{3}} - \frac{d \phi^{\star}}{dx}\frac{d^{2} \varphi}{dx^{2}} + \frac{d \varphi}{dx}\frac{d^{2} \phi^{\star}}{dx^{2}} \right]^{b}_{a} = 0.
\end{equation}

In the finite square-well potential case is easy verify that\footnote{It is already familiar result that the representation (\ref{xrep}) and (\ref{prep}) is Hermitian \cite{Fring}.} $\hat{H} = \hat{H}^{\dag}$ because of the boundary conditions, Eqs. (\ref{bc1}) and (\ref{bc2}). Consequently, the $\hat{U}$ time evolution operator is unitary.

Meanwhile, some caution is necessary with the infinite square-well potential case. If $\varphi, \phi \in D(\hat{H}) = \left\{ \varphi, \hat{H}\varphi \in {\cal L}^{2} \left(-\frac{a}{2}, +\frac{a}{2} \right), \varphi \left( \pm \frac{a}{2} \right) = \varphi^{\prime} \left( \pm \frac{a}{2} \right) = 0 \right\}$ then $\hat{H}$ is symmetric, but it is not self-adjoint\footnote{Apparently, the deficiency indices are equal ($n_{+} = n_{-}$). On the superficial analysis, probably they are $(4,4)$ and $\hat{H}$, with $D_{0}(\hat{H}) = \left\{ \varphi, \hat{H}\varphi \in {\cal L}^{2} \left(-\frac{a}{2}, +\frac{a}{2} \right),  \varphi^{(n)} \left( \pm \frac{a}{2} \right) = 0, n = 0,1,2,3 \right\}$, has infinitely many self-adjoint extensions, a $U(4)$ family of self-adjoint extensions. Nevertheless, the determination of the deficiency indices and of the possible self-adjoint extensions can be a no easy task. The case of  a $\hat{H}$ that has a $\hat{p}_{0}^{3}$ correction term has been studied by J. Louko \cite{Louko}.}. This is not surprise since something similar takes place in the ordinary quantum mechanics \cite{Bonneau,Schechter,Karwowski,Araujo}. Note that if we only impose that $\varphi(x)= 0$ at $x = \pm \frac{a}{2}$ it will not be enough to became $\hat{H}$ symmetric.



\section{Conclusion}
\label{Concl}

\h In this work we solve, in a minimal-length scenario, the problem of a finite square-well potential in 1-dim, whose boundary conditions are well fixed. We then take the limit of the potential going to infinity in order to determine the solutions for an infinite square-well potential. In contrast to what has been found in the literature, the eigenfunctions of a particle within an infinite square-well in a minimal-length scenario are not the same as those in ordinary quantum mechanics and it is not necessary to use the energy equation to ensure the continuity of the eigenfunctions at the walls of the well, but the use them ensures the continuity of the first derivatives of the eigenfunctions at the walls.

It is important to emphasize that, as we have seen, the continuity of the first derivative is necessary condition to guarantee the continuity of the probability current density at $x = \pm \frac{a}{2}$  and also to ensure the unitarity of the time evolution operator.

In summary, in order to solve an infinite square-well potential in a minimal-length scenario we must require the continuity of the wave function and its first derivative, that is, they vanish at the walls of the infinite square-well. Therefore, there are four coefficients to determine and four equations obtained from the boundary conditions. However, three of the equations allow to find three coefficients as function of a fourth coefficient (which is determined by the normalization condition) while the fourth equation allows to find the energy equation.

Finally, nowadays there is a huge amount of works on trapped ions and particles in which the trap system is modelled by quantum wells. At first, such systems could be used to provide an upper bound for the minimal-length value. However, one must be very careful with the result obtained since it is very likely that the error in modelling a trap of an electronic devise by a quantum well is bigger than the minimal-length correction \cite{blado}. So, this must be very verified for each experimental set up employed.


\section*{Acknowledgements}

\h We would like to thank FAPES, CAPES and CNPq (Brazil) for financial support.
\\


\appendix

\section{Solution of the System of Linear Equations}
\label{sle}
\noindent

Making use that the potential is even and imposing the boundary conditions
$$
\frac{d^{n} \varphi_{II} (x)}{x^{n}} \bigg|_{x = \frac{a}{2}}  = \frac{d^{n} \varphi_{III}(x)}{x^{n}} \bigg|_{x = \frac{a}{2}},
$$
for $n = 0, 1, 2, 3$, we obtain the system of equations (A1-A4),
\begin{equation*}
A_{2} \cos \left(\frac{k_{0}^{\beta} a}{2} \right) + C_{2} \cosh \left(\frac{k_{\beta}^{+} a}{2} \right) = 
\end{equation*}
\begin{equation}
\label{a1}
A_{3} \exp {\left( -\frac{k_{1}^{\beta} a}{2} \right)} + C_{3} \exp {\left( -\frac{k_{\beta}^{-} a}{2} \right)}
\end{equation}
\\
\begin{equation*}
-k_{0}^{\beta}A_{2} \sin \left(\frac{k_{0}^{\beta} a}{2} \right) + k_{\beta}^{+} C_{2} \sinh \left(\frac{k_{\beta}^{+} a}{2} \right) =
\end{equation*}
\begin{equation}
\label{a2}
-k_{1}^{\beta} A_{3} \exp {\left( -\frac{k_{1}^{\beta} a}{2} \right)} - k_{\beta}^{-} C_{3} \exp {\left( -\frac{k_{\beta}^{-} a}{2} \right)}
\end{equation}
\\
\begin{equation*}
-\left( k_{0}^{\beta} \right)^{2} A_{2} \cos \left(\frac{k_{0}^{\beta} a}{2} \right) + \left( k_{\beta}^{+} \right)^{2} C_{2} \cosh \left(\frac{k_{\beta}^{+} a}{2} \right) =
\end{equation*}
\begin{equation}
\label{a3}
\left( k_{1}^{\beta} \right)^{2} A_{3} \exp {\left( -\frac{k_{1}^{\beta} a}{2} \right)} + \left( k_{\beta}^{-} \right)^{2} C_{3} A_{3} \exp {\left( -\frac{k_{\beta}^{-} a}{2} \right)}
\end{equation}
\\
\begin{equation*}
\left( k_{0}^{\beta} \right)^{3} A_{2} \sin \left(\frac{k_{0}^{\beta} a}{2} \right) + \left( k_{\beta}^{+} \right)^{3} C_{2} \sinh \left(\frac{k_{\beta}^{+} a}{2} \right) = 
\end{equation*}
\begin{equation}
\label{a4}
-\left( k_{1}^{\beta} \right)^{3} A_{3} \exp {\left( -\frac{k_{1}^{\beta} a}{2} \right)} -  \left( k_{\beta}^{-} \right)^{3} C_{3} \exp {\left( -\frac{k_{\beta}^{-} a}{2} \right)},
\end{equation}
which, along with normalization condition, enables us to determine the coefficients of the even solutions.

Multiplying Eq. (\ref{a1}) by $\left( k_{\beta}^{-} \right)^{2}$ and subtracting from the result Eq. (\ref{a3}) we have
\begin{equation*}
\left[ \left(k_{\beta}^{-} \right)^{2} + \left(k_{0}^{\beta} \right)^{2} \right] A_{2} \cos \left(\frac{k_{0}^{\beta} a}{2} \right) + \left[ \left(k_{\beta}^{-} \right)^{2} - \left(k_{\beta}^{+} \right)^{2} \right] C_{2} \cosh \left(\frac{k_{\beta}^{+} a}{2} \right) =
\end{equation*}
\begin{equation}
\label{a5}
\left[ \left(k_{\beta}^{-} \right)^{2} - \left(k_{1}^{\beta} \right)^{2} \right] A_{3} \exp {\left( -\frac{k_{1}^{\beta} a}{2} \right)}.
\end{equation}
Now, multiplying Eq. (\ref{a2}) by $-\left( k_{\beta}^{-} \right)^{2}$ and adding the result to Eq. (\ref{a4}) we obtain
\begin{equation*}
k_{0}^{\beta}\left[ \left(k_{\beta}^{-} \right)^{2} + \left(k_{0}^{\beta} \right)^{2} \right] A_{2} \sin \left(\frac{k_{0}^{\beta} a}{2} \right) - k_{\beta}^{+} \left[ \left(k_{\beta}^{-} \right)^{2} - \left(k_{\beta}^{+} \right)^{2} \right] C_{2} \sinh \left(\frac{k_{\beta}^{+} a}{2} \right) =
\end{equation*}
\begin{equation}
\label{a6}
 k_{1}^{\beta}\left[ \left(k_{\beta}^{-} \right)^{2} + \left(k_{1}^{\beta} \right)^{2} \right] A_{3} \exp {\left( -\frac{k_{1}^{\beta} a}{2} \right)}.
\end{equation}
In order to find the coefficient $C_{2}$ as a function of $A_{2}$ we multiply Eq. (\ref{a5}) by $k_{1}^{\beta}$ and subtract from result Eq. (\ref{a6}). Hence,
\begin{equation}
\label{a7}
C_{2} = - \frac{ \left( k^{-}_{\beta} \right)^{2} + \left( k_{0}^{\beta} \right)^{2}}{\left( k^{-}_{\beta} \right)^{2}  - \left( k_{\beta}^{+} \right)^{2}} \left[ \frac{ k_{1}^{\beta} - k_{0}^{\beta}\tan \left( \frac{ k_{0}^{\beta} a}{2} \right)}{ k_{1}^{\beta} + k_{\beta}^{+} \tanh \left( \frac{ k_{\beta}^{+} a}{2} \right)} \right] \frac{\cos \left( \frac{ k_{0}^{\beta} a}{2} \right)} {\cosh \left( \frac{ k_{\beta}^{+} a}{2} \right)} A_{2}.
\end{equation}
Now, using the result (\ref{a7}) into (\ref{a5}) we obtain the coeffcient $A_{3}$ as a function of $A_{2}$,
\begin{equation}
\label{a8}
A_{3} = \frac{ \left( k^{-}_{\beta} \right)^{2} + \left( k_{0}^{\beta} \right)^{2}}{\left( k^{-}_{\beta} \right)^{2}  - \left( k_{1}^{\beta} \right)^{2}} \left[ \frac{ k_{0}^{\beta}\tan \left( \frac{ k_{0}^{\beta} a}{2} \right) + k_{\beta}^{+}\tanh \left( \frac{ k_{\beta}^{+} a}{2} \right)}{ k_{1}^{\beta} + k_{\beta}^{+}\tanh \left( \frac{ k_{\beta}^{+} a}{2} \right)} \right] \cos \left( \frac{ k_{0}^{\beta} a}{2} \right) \exp \left( \frac{k_{1}^{\beta} a}{2} \right) A_{2}.
\end{equation}
At last, using the results (\ref{a7}) and (\ref{a8}) into Eq. (\ref{a1}) we get the coefficient $C_{3}$ as a function of $A_{2}$,
\begin{equation*}
C_{3} = \Bigg\{ 1 - \frac{ \left( k^{-}_{\beta} \right)^{2} + \left( k_{0}^{\beta} \right)^{2}}{\left( k^{-}_{\beta} \right)^{2}  - \left( k_{1}^{\beta} \right)^{2}} \left[ \frac{ k_{0}^{\beta}\tan \left( \frac{ k_{0}^{\beta} a}{2} \right) + k_{\beta}^{+}\tanh \left( \frac{ k_{\beta}^{+} a}{2} \right)}{ k_{1}^{\beta} + k_{\beta}^{+}\tanh \left( \frac{ k_{\beta}^{+} a}{2} \right)} \right] -
\end{equation*}
\begin{equation}
\label{a9}
\frac{ \left( k^{-}_{\beta} \right)^{2} + \left( k_{0}^{\beta} \right)^{2}}{\left( k^{-}_{\beta} \right)^{2}  - \left( k_{\beta}^{+} \right)^{2}} \left[ \frac{ k_{1}^{\beta} - k_{0}^{\beta}\tan \left( \frac{ k_{0}^{\beta} a}{2} \right)}{ k_{1}^{\beta} + k_{\beta}^{+} \tanh \left( \frac{ k_{\beta}^{+} a}{2} \right)} \right] \Bigg\} \cos \left( \frac{ k_{0}^{\beta} a}{2} \right) \exp \left( \frac{k_{\beta}^{-}a}{2} \right) A_{2}.
\end{equation}

The use of the rsults (\ref{a7}), (\ref{a8}) and (\ref{a9}) does not allow us to determine the coefficient $A_{2}$ unically. Instead of this, we obtain an equation which allows the energy of the system only takes on certain values,
\begin{equation*}
k_{0}^{\beta}\tan \left( \frac{ k_{0}^{\beta} a}{2} \right) - k_{\beta}^{-} +
\end{equation*}
\begin{equation*}
\frac{ \left( k^{-}_{\beta} \right)^{2} + \left( k_{0}^{\beta} \right)^{2}}{\left( k^{-}_{\beta} \right)^{2}  - \left( k_{\beta}^{+} \right)^{2}}\left[ \frac{ k_{1}^{\beta} - k_{0}^{\beta}\tan \left( \frac{ k_{0}^{\beta} a}{2} \right)}{ k_{1}^{\beta} + k_{\beta}^{+} \tanh \left( \frac{ k_{\beta}^{+} a}{2} \right)} \right]\left[k_{\beta}^{-} + k_{\beta}^{+} \tanh \left( \frac{ k_{\beta}^{+} a}{2} \right) \right] +
\end{equation*}
\begin{equation}
\label{a10}
\frac{ \left( k^{-}_{\beta} \right)^{2} + \left( k_{0}^{\beta} \right)^{2}}{\left( k^{-}_{\beta} \right)^{2}  - \left( k_{1}^{\beta} \right)^{2}}\left[ \frac{ k_{0}^{\beta}\tan \left( \frac{ k_{0}^{\beta} a}{2} \right) + k_{\beta}^{+}\tanh \left( \frac{ k_{\beta}^{+} a}{2} \right)}{ k_{1}^{\beta} + k_{\beta}^{+}\tanh \left( \frac{ k_{\beta}^{+} a}{2} \right)} \right]\left( k_{\beta}^{-} - k_{1}^{\beta} \right) = 0.
\end{equation}

As it is well-known, the coefficient $A_{2}$ can be determined unically by the normalization condition, however its determination is laborious and  not important for the goals of this work.

Just as we expect, in the limit when $\beta$ goes to zero ($\beta \rightarrow 0$) the solution (\ref{esol}) becomes\footnote{Note that the second terms of Eq. (\ref{esol}) vanish because $e^{\frac{1}{\sqrt{\beta}}}$ goes to infinity faster than $\frac{1}{\beta}$ when $\beta \rightarrow 0$.}
\begin{equation}
\label{esolb0} 
	\left \{
	\begin{array}{ll}
	\varphi_{I}(x) = A_{2} \cos \left(\frac{k_{0} a}{2} \right) e^{k_{1} \left(x + \frac{a}{2} \right)}  & x \leq -\frac{a}{2} \\ [11pt]
	\varphi_{II}(x) = A_{2}\cos \left( k_{0}^{\beta} x \right), & -\frac{a}{2} \leq x \leq \frac{a}{2}, \\ [11pt]
	\varphi_{III}(x) = A_{2} \cos \left(\frac{k_{0} a}{2} \right) e^{- k_{1} \left(x - \frac{a}{2} \right)} & x \geq \frac{a}{2} \\
	\end{array}
	\right.
\end{equation}
and the energy equation (\ref{eeeq}) becomes
\begin{equation}
\label{eeeqb0}
\tan \left( \frac{k_{0} a}{2} \right) = \frac{k_{1}}{k_{0}},
\end{equation}
which are the even solutions and the energy equation for the finite square-well potential in the ordinary case $\beta = 0$.

The odd solutions can be obtained in the similar way.





\end{document}